\documentclass[final,5p,times,twocolumn]{elsarticle}

\usepackage{amssymb}
\usepackage{amsmath}
\journal{Physics letters B}

\begin{document}

\begin{frontmatter}



\title{General Expressions for On-shell Recursion relations for Tree-level Open String Amplitudes}


\author{Pongwit Srisangyingcharoen}
\ead{pongwits@nu.ac.th}
\affiliation{organization={The Institute for Fundamental Study, Naresuan University},%Department and Organization
            city={Phisanulok},
            postcode={65000}, 
            country={Thailand}}

\begin{abstract}
In this paper, we present a systematic derivation aimed at obtaining general expressions for on-shell recursion relations for tree-level open string amplitudes. Our approach involves applying the BCFW shift to an open string amplitude written in terms of multiple Gaussian hypergeometric functions. By employing binomial expansions, we demonstrate that the shifted amplitudes manifest simple poles, which correspond to scattering channels of intermediate states. Using the residue theorem, we thereby derive a general expression for these relations.
\end{abstract}



\begin{keyword}
Bosonic string, string amplitudes, on-shell recursion relations


\end{keyword}

\end{frontmatter}


\section{Introduction}\label{sec1}
Since the beginning of the string theory, understanding string scattering amplitudes has been a fundamental concern among string theorists. It was long known that in the limit of low energies, amplitudes in string theory reproduce those in QFT such as Yang-Mills \cite{Neveu:1971mu} and Einstein theory \cite{Scherk:1974ca, Yoneya:1974jg} plus $\alpha'$ corrections \cite{TSEYTLIN1986391, Koerber:2001uu, Metsaev:1986yb, Bergshoeff:1989de, Garousi:2019mca}. The connection between string theory and QFT provides useful applications in both theories. Understanding the structure of string amplitudes would provide a better insight into those of quantum field theories. Concrete examples are the celebrated Kawai-Lewellen-Tye (KLT) relations \cite{Kawai:1985xq} which relates closed string amplitudes in terms of products of two open string amplitudes giving alternative descriptions of gravity as the square of gauge theory. These non-linear relations were proven later in the context of QFT \cite{Bjerrum-Bohr:2010kyi, Bjerrum-Bohr:2010mtb}.

 Another interesting structure was discovered by Plahte \cite{Plahte:1970wy} which are linear relations among color-ordered open string scattering amplitudes. These are currently known as monodromy relations. In the field theory limit, the relations reduce to the BCJ relations of Bern, Carrasco and Johansson \cite{Bern:2008qj} and the Kleiss-Kujif relations \cite{Kleiss:1988ne}. This results in a reduction of the number of color-ordered amplitudes from $(n-1)!$ as given by a cyclic property of the trace down to $(n- 3)!$ \cite{Bjerrum-Bohr:2009ulz, Stieberger:2009hq}. The monodromy relations among partial open string amplitudes can be captured by polygons in the complex plane \cite{Srisangyingcharoen:2020lhx}.

 %%%%%%%%%%%%%%%%%%%%%%%%%%%% Review BCFW (needed to paraphrase or rewrite)
During the early 2000s, advancements in the study of scattering amplitudes were notably influenced by the discovery of the Britto-Cachazo-Feng-Witten (BCFW) on-shell recursion relations \cite{Britto_2005,Britto2005NewRR}. These relations enabled the expression of tree-level amplitudes as products involving amplitudes with fewer particles. The key idea for deriving the on-shell recursion relations is based on the fact that any tree-level scattering amplitude is a rational function of the external momenta, thus, one can turn an amplitude $A_n$ into a complex meromorphic function $A_n(z)$ by deforming the external momenta through introducing a complex variable $z$. These deformed momenta, satisfying momentum conservation, are required to remain on-shell. For a scattering process involving $n$ particles, the selection of an arbitrary pair of particle momenta for shifting is permissible. Our choice is given by
\begin{subequations} \label{BCFW shift}
\begin{align}
    k_1 \rightarrow \hat{k}_1(z) =& k_1-qz \\
    k_n \rightarrow \hat{k}_n(z) =& k_n+qz
\end{align}
\label{shifted momenta2}
\end{subequations}
where $q$ is a reference momentum which obeys $q\cdot q=k_1\cdot q=k_n\cdot q=0$. 

The unshifted amplitude $A_n(z=0)$ can be obtained from a contour integration in which the contour is large enough to enclose all finite poles. According to the Cauchy's theorem,
\begin{equation}
    A_n(0)=\oint dz \frac{A_n(z)}{z}-\sum_\text{poles} \text{Res}_{z=z_\text{poles}}\left(  \frac{A_n(z)}{z} \right), \label{cauchy}
\end{equation}
the unshifted amplitude at $z=0$ is equal to the sum of the residues over all the finite poles if the amplitude is well-behaved at large $z$ (which is the case for  most theories). For Yang-Mills theory, the residue at a finite pole is the product of two fewer-point amplitudes with an on-shell exchanged particle. In Yang-Mills a sum over the helicities of the intermediate gauge boson and in general theories a sum over all allowed intermediate particle states must also be done. In the general case, the BCFW recursion relation is 
\begin{equation}
    A_n(0)=\sum_{\substack{\text{poles}\\ \alpha}}\sum_{\substack{\text{physical}\\ \text{states}}} A_L(\dots,P(z_\alpha))\frac{2}{P^2+M^2}A_R(-P(z_\alpha),\dots)
    \label{BCFW}
\end{equation}
with $P$ being the momentum of the exchanged particle with mass $M$. 

The validity of equation (\ref{cauchy}) requires the absence of a pole at infinity. In the case that there exists such a pole, one must include the residue at infinity. However, the residue at this pole does not have a similar physical interpretation to the residues at finite poles. A detailed discussion can be found in \cite{Feng_2010}.

 %%%%%%%%%%%%%%%%%%%%%%% Objectives and outlines of the paper
This paper aims to present expressions for on-shell recursion relations concerning open string amplitudes. While existing literature has explored relations for open strings \cite{RutgerBoels_2008, Boels:2010bv, Chang:2012qs, Cheung:2010vn}, general forms for open string on-shell recursion relations have never been delivered. Our paper aims to present a systematic derivation to such relations based on the Koba-Nielsen integral forms \cite{Koba:1969rw}.

\section{On-shell Recursion Relations for Four-point Tachyonic Open String Amplitudes}\label{sec2}
Before deriving a general expression for the on-shell recursion relations of open string amplitudes, let's start with a discussion of the simplest example, i.e. the four-point tachyon amplitude. Consider the partial amplitude
\begin{align}
    \mathcal{A}(s,t)=\int_0^1 dx \ x^{\alpha's-2}(1-x)^{\alpha't-2}
\end{align}
where $s$ and $t$ are the Mandelstam variables given by $s=(k_1+k_2)^2$ and $t=(k_1+k_4)^2$. The open string vertex variables $x_1, x_3$ and $x_4$ are fixed to $0, 1$ and $\infty$ due to the $PSL(2,R)$ gauge symmetry. Under the shift in (\ref{BCFW shift}), $\mathcal{A}(s,t)\rightarrow \mathcal{A}(z')$ giving
\begin{align}
    \mathcal{A}(z')=\int_0^1 dx \  x^{\alpha's-2-z'}(1-x)^{\alpha't-2} \label{A4-1}
\end{align}
where $z'=2\alpha' z q\cdot k_2$. To determine poles generated from $z'$, we introduce the variable 
\begin{equation}
    x=\exp{\left( -\frac{y}{\alpha's-2-z'} \right)} \qquad \text{for} \quad \text{Re}(\alpha's-2-z')>0. \label{coor transf}
\end{equation}
The integral (\ref{A4-1}) becomes
\begin{align}
    \mathcal{A}(z')=\sum_{a=0}^\infty \binom{\alpha't-2}{a}(-1)^a \int_0^\infty dy \ e^{-y} \frac{e^{-\frac{(1+a)y}{\alpha's-2-z'}}}{\alpha's-2-z'} \label{deformed 4-point}
\end{align}
where a binomial expansion $(1-x)^a=\sum_{k=0}^\infty\binom{a}{k}(-1)^k x^k$ was applied. $\binom{a}{k}$ denotes a binomial coefficient. Expanding a Taylor series for the exponential, one obtains
\begin{align}
       \mathcal{A}(z')&=\sum_{a=0}^\infty \binom{\alpha't-2}{a}(-1)^a \sum_{k=0}^\infty \frac{(-1)^k}{k!}\int_0^\infty dy \ e^{-y} \frac{((1+a)y)^k}{(\alpha's-2-z')^{k+1}}.\label{deformed 4-point2}  \\
       &=-\sum_{a=0}^\infty \binom{\alpha't-2}{a}(-1)^a \sum_{k=0}^\infty  \frac{(1+a)^k}{(z'-(\alpha's-2))^{k+1}} \label{deformed 4-point3}
\end{align}
This shows that $\mathcal{A}(z')$ contains the $k^{\text{th}}$-order poles at $z'=\Tilde{z}\equiv\alpha's-2$.
We can regain the unshifted amplitude $\mathcal{A}(0)$ using the relation (\ref{cauchy}). Therefore,
\begin{align}
    \mathcal{A}(0)&=-\text{Res}_{z'=\Tilde{z}}\left(  \frac{\mathcal{A}(z')}{z'} \right)=\sum_{a,k=0}^\infty \binom{\alpha't-2}{a}\frac{(-1)^{a+k}}{\Tilde{z}}\left(\frac{1+a}{\Tilde{z}}\right)^k \nonumber \\
    &=\sum_{a=0}^\infty \binom{\alpha't-2}{a} \frac{(-1)^a}{\alpha's-(1-a)}. \label{tach 4point}
\end{align}
The expression implies propagators of intermediate on-shell string states. Comparing to the BCFW recursion relation (\ref{BCFW}), the residues of (\ref{tach 4point}) are sum over product of physical state amplitudes at each fixed level $a$. This was explicit shown in \cite{Chang:2012qs}.

It is worth noting that the derivation of the expression (\ref{tach 4point}) assumes the 
condition such that Re$(\alpha's-2-z')>0$ during the coordinates transformation (\ref{coor transf}). To obtain the same argument but for the kinematic regime Re$(\alpha's-2-z')<0$ is a bit trickier as we need to deal with divergence which requires a proper regularization. To see this, for Re$(\alpha's-2-z')<0$, we use a change of variables
\begin{equation}
     x=\exp{\left( \frac{y}{\alpha's-2-z'} \right)} \label{4pt new var2}
\end{equation}
to turn the amplitude (\ref{deformed 4-point}) into
\begin{align}
    \mathcal{A}(z')&=\sum_{a=0}^\infty \binom{\alpha't-2}{a}(-1)^{a+1} \sum_{k=0}^\infty \frac{1}{k!}\int_0^\infty dy \ e^{y} \frac{((1+a)y)^k}{(\alpha's-2-z')^{k+1}} \label{tach 4point2}
\end{align}
where the binomial expansion was applied. Notice a slight difference from (\ref{deformed 4-point2}) due to the minus signs. Now comes the divergence integral $\int_0^\infty dy \ e^{y} y^{s}$ by which we can regularize it to be 
\begin{equation}
    \int_0^\infty dy \  y^{s}e^{y}\coloneqq (-1)^{s+1} s! \qquad \text{for} \quad s \in \mathbb{Z}^+\cup\{0\}. \label{reg integral}
\end{equation}
The symbol $\coloneqq$ signifies that the equality holds upon the regularization. In this case, we assign the value for the integral via analytic continuation of the parameter $n$ from
\begin{equation}
    \frac{1}{A^n}=\frac{1}{(n+1)!}\int_0^\infty dx \ x^{n-1}e^{-xA}.
\end{equation}
 Accordingly, substituting (\ref{reg integral}) into (\ref{tach 4point2}), one would regain the same expression of $\mathcal{A}(z')$ as (\ref{deformed 4-point3}), hence providing the same on-shell recursion relation (\ref{tach 4point}).

\section{On-shell Recursion Relations for $n-$point Open String Amplitudes}\label{sec3}

To generalize the investigation to a general $n$-point open string amplitude, we consider a general expression 
\begin{align}
    \mathcal{A}(1,2,3,
    \ldots,n)=\int_{\mathcal{I}}\prod_{i=1}^n dz_i \frac{\vert z_{ab}z_{ac}z_{bc} \vert}{dz_a dz_b dz_c} \prod_{1\leq i<j\leq n}\vert x_i-x_j\vert^{2\alpha'k_i\cdot k_j}\mathcal{F}_n \label{Koba-Nielsen n-point}
\end{align}
where $z_{ij}=x_i-x_j$ and $dz_i=dx_i$ for bosonic string theory and $z_{ij}=x_i-x_j+\theta_i\theta_j$ and $dz_i=dx_i d\theta_i$ for the supersymmetric case. The integral is subject to the integration region $\mathcal{I}$ where $x_1<x_2 \ldots <x_n$ which is associated to the group factor tr($T_1T_2\ldots T_n$). The gauge symmetry $PSL(2,\mathbb{R})$ allow one to fix the position of three points denoted $z_a, z_b$ and $z_c$. A conventional choice is $x_1=0, x_{n-1}=1$ and $x_n=\infty$ for the bosonic string as well as $\theta_{n-1}=\theta_n=0$ for for the supersymmetric case.

The function $\mathcal{F}_n$ is a branch-free function that comes from the operator product expansion of vertex operators depending on the external states of the amplitude we consider. $\mathcal{F}_n=1$ for tachyons and $\mathcal{F}_n=\text{exp}\Big( \sum_{i>j}\frac{\xi_i\cdot \xi_j}{(x_i-x_j)^2}-\sum_{i\neq j}\frac{\sqrt{\alpha'}k_i\cdot \xi_j}{(x_i-x_j)}  \Big)\Big\vert_\text{multilinear in $\xi_i$}$ for an $n$-gauge field amplitude with $n$ polarization vectors $\xi_i$. In addition, $\mathcal{F}_n=\int\prod_{i=1}^{n}d\eta_i \allowbreak \times \text{exp}\Big[\sum_{i\neq j} \Big(  \frac{\sqrt{\alpha'}\eta_i(\theta_i-\theta_j)(\xi_i\cdot k_j)-\eta_i\eta_j(\xi_i\cot \xi_j)}{(x_i-x_j+\theta_i\theta_j)}\Big) \Big]$ for the superstring amplitude where $\eta_i$ are Grassmann variables.

Alternatively, one could relate the Koba-Nielsen's integral representation of open string amplitudes (\ref{Koba-Nielsen n-point}) with multiple Gaussian hypergeometric functions,
\begin{align}
    B_n(\{ n\})=\left(\prod_{i=1}^{n-3}\int_0^1 dw_i \right) \prod_{j=1}^{n-3} w_j^{s_{12\ldots j+1}+n_j} \prod_{l=j}^{n-3}\left( 1-\prod_{k=j}^l w_k \right)^{s_{j+1,l+2}+n_{j+1,l+2}} \label{hypergeometric}
\end{align}
where $s_{ij}=\alpha'(k_i+k_j)^2$ and $s_{12\ldots i}=\alpha'(k_1+k_2+\ldots+k_i)^2$. The set $\{n\}$ contains all integers $n_i$ and $n_{ij}$ appearing on the right-hand side of (\ref{hypergeometric}). More precisely, the function (\ref{hypergeometric}) is known as generalized Kamp\'e de F\'eriet function \cite{Srivastava1985MultipleGH}. To obtain the above expression, we applied a change of integral variables 
\begin{equation}
    x_i=\prod_{j=i-1}^{n-3} w_j \qquad \text{for} \quad i=2,3,\ldots, n-2
\end{equation}
to
\begin{equation}
    \int_{\mathcal{I}}\prod_{i=2}^{n-2}dx_i \ \prod_{1\leq i<j\leq n}\vert x_i-x_j\vert^{2\alpha'k_i\cdot k_j+\tilde{n}_{ij}}.
\end{equation}
Note that we have fixed $x_1, x_{n-1}$ and $x_n$ to $0, 1$ and $\infty$ respectively. The exponents $\tilde{n}_{ij}\in \mathbb{Z}$ which arise from the external state-dependent term $\mathcal{F}_n$. They relate to the integers $n_i$ and $n_{ij}$ via
\begin{align}
    n_{ij}&=\tilde{n}_{ij}+\alpha'(m_{i}^2+m_{j}^2), && i\leq j \nonumber \\
    n_j&=j-1+\sum_{l=1}^{j}\sum_{k>l}^{j+1}\tilde{n}_{lk}+\alpha'\sum_{i=1}^{j+1}m_i^2, && 1\leq j\leq n-3. \label{n n}
\end{align}
More detail can be found in Chapter 7 of \cite{https://doi.org/10.1002/prop.201100084}. Consequently, the general expression of color-ordered open string amplitudes reads
\begin{equation}
    \mathcal{A}(1,2,\ldots,n)=\sum_I \mathcal{K}_I B_n(\{n^I\}) \label{n-point amp}
\end{equation}
where the function $\mathcal{K}_I$ contains scalar products of momentum and polarization vectors which are $k_i\cdot k_j, k_i\cdot \xi_j$ and $\xi_i\cdot \xi_j$ \cite{Stieberger:2006te}.

Rewriting the open string amplitude as (\ref{n-point amp}) is useful to determine the pole structure of the deformed amplitude $\mathcal{A}_n(z)$. This allows us to formulate the on-shell recursion relations for $n$-point open string amplitudes. When the momenta are shifted using (\ref{BCFW shift}), the function $B_n(\{ n\})$ becomes
\begin{align}
    B_n(\{ n\},z)=\left(\prod_{i=1}^{n-3}\int_0^1 dw_i \right) &\prod_{j=1}^{n-3} w_j^{s_{12\ldots j+1}+n_j-z_{j}} \nonumber \\
    &\times \prod_{l=j}^{n-3}\left( 1-\prod_{k=j}^l w_k \right)^{s_{j+1,l+2}+n_{j+1,l+2}} \label{deformed hypergeometric}
\end{align}
where $z_i=2\alpha'z (q\cdot \sum_{j=2}^{i+1} k_j)$. This implies that there are $n-3$ poles generated from $z_i$ which refers to $n-3$ scattering channels. To determine the poles of $z_i$, we apply binomial expansions to every term
that contains $w_i$ in the product
\begin{equation}
   \prod_{j=1}^{n-3} \prod_{l=j}^{n-3}\left( 1-\prod_{k=j}^l w_k \right)^{s_{j+1,l+2}+n_{j+1,l+2}}. \label{product polynomial}
\end{equation}

For example, let's first start by determining the pole regarding $z_1$. Therefore, one needs to rewrite
\begin{align}
    \prod_{l=1}^{n-3}\left( 1-\prod_{k=1}^l w_k \right)^{s_{2,l+2}+n_{2,l+2}}= \prod_{l=1}^{n-3}\sum_{a_l=0}^\infty \binom{s_{2,l+2}+n_{2,l+2}}{a_l}(-1)^{a_l}w_l^{\sum_{i=l}^{n-3}a_i}
\end{align}
using the binomial expansion. The expression involves $n-3$ parameters $a_l$ resulted from the expansions. This turns (\ref{deformed hypergeometric}) to be
\begin{align}
    B_n(\{ n\},z)=& \prod_{l=1}^{n-3}\bigg(\sum_{a_l=0}^\infty \binom{s_{2,l+2}+n_{2,l+2}}{a_l} \bigg)(-1)^{\sum_{i=1}^{n-3}a_i} \nonumber \\
    &\times \bigg[\int_0^1 dw_1 w_1^{s_{12}+n_1+\sum_{i=1}^{n-3}a_i-z_1}\bigg]\prod_{i=2}^{n-3}\int_0^1 dw_i \nonumber \\
    &\times \prod_{j=2}^{n-3}w_j^{s_{12\ldots j+1}+n_j+\sum_{i=j}^{n-3}a_i-z_j}\prod_{l=2}^{n-3}\left( 1-\prod_{k=2}^l w_k \right)^{s_{3,l+2}+n_{3,l+2}}. \label{B1}
\end{align}
We then rename the sum over all $a_l$ to the new parameter, says
\begin{equation}
    \sum_{i=1}^{n-3}a_i=m
\end{equation}
which runs from zero to infinity. This index $m$ labels the poles from $z_1$. Notice that the integral in the square bracket of (\ref{B1}) in the second line produces the simple poles characterized by the index $m$
\begin{equation}
    \frac{(-1)}{z_1-(s_{12}+n_1+m+1)}.
\end{equation}
As a result, one can compute the residue 
\begin{align}
    \sum_m \text{Res}_{z=z^*_m}\bigg(\frac{B_n(\{ n\},z)}{z}\bigg)=\sum_m \frac{\mathcal{B}_1(\{n\},m,z^*_m)}{s_{12}+n_1+m+1} \label{residue z1}
\end{align}
where $z^*_m=(s_{12}+n_1+m-1)/(2\alpha' q\cdot k_2)$. The function $\mathcal{B}_1(\{n\},m,z)$ is defined as
\begin{align}
    \prod_{l=1}^{n-4}\left(\sum_{a_l=0}^\infty \binom{s_{2,l+2}+n_{2,l+2}}{a_l} \right)\binom{s_{2,n-1}+n_{2,n-1}}{m-\sum_{l=1}^{n-4}a_l}(-1)^{m+1}\prod_{i=2}^{n-3}\int_0^1 dw_i \nonumber \\
    \times \prod_{j=2}^{n-3}w_j^{s_{12\ldots j+1}+n_j+m-\sum_{i=1}^{j-1}a_i-z_j}\prod_{l=2}^{n-3}\left( 1-\prod_{k=2}^l w_k \right)^{s_{3,l+2}+n_{3,l+2}}. \label{B1-2}
\end{align}
Note that $\mathcal{B}_1[n_i,n_{ij},z]$ is evaluated at $z=z^*_m$ in (\ref{residue z1}). The denominators of the equation (\ref{residue z1}) suggest the mass spectrum of intermediate states with momentum $k_1+k_2$. What we need to do next is to generalize this kind of calculation to involve all poles corresponding to all $z_j$.

Similar to the previous calculation, to determine the poles associated with $z_i$ for $i=1,2,\ldots,n-3$, one requires a binomial expansion to (\ref{product polynomial}) to expand all the terms that have $w_i$. These expansions would generate $(n-2-i)i$ summing indices. As for the previous case, we had $n-3$ indices for $i=1$. 

For convenience, we will assign the summing index $a_{jl}$ for $j\leq l$ when expanding the polynomial
\begin{align}
    \left( 1-\prod_{k=j}^l w_k \right)^{\tilde{s}_{j+1,l+2}}=\sum_{a_{jl}=0}^\infty \binom{\tilde{s}_{j+1,l+2}}{a_{jl}}(-1)^{a_{jl}} \left(\prod_{k=j}^l w_k\right)^{a_{jl}} \label{binomial exp}
\end{align}
where $\tilde{s}_{j+1,l+2}\equiv s_{j+1,l+2}+n_{j+1,l+2}$ for short. We will use the above expansion to the product (\ref{product polynomial}) for which $j\leq i\leq l$ (Note that $i$ refers to the index where the poles $z_i$ will be determined). This gives
\begin{align}
       \prod_{j\leq l}&\left( 1-\prod_{k=j}^l w_k \right)^{\tilde{s}_{j+1,l+2}}= \nonumber \\
       &\prod_{\substack{j\leq l\\i \notin [j,l]}}\left( 1-\prod_{k=j}^l w_k \right)^{\tilde{s}_{j+1,l+2}} \prod_{\substack{j\leq l\\i \in [j,l]}}\left(  \sum_{a_{jl}=0}^\infty \binom{\tilde{s}_{j+1,l+2}}{a_{jl}} (-1)^{a_{jl}} \left(\prod_{k=j}^l w_k  \right)^{a_{jl}} \right).
\end{align}
Note that we only apply the expansions for the case where $i \in [j,l]$ as discussed. Accordingly, we can write (\ref{deformed hypergeometric}) as
\begin{align}
    B_n(\{ n\},z)=&\left(\prod_{\substack{k=1\\k\neq i}}^{n-3}\int_0^1 dw_k \right) \prod_{\substack{j=1\\j\neq i}}^{n-3} w_j^{s_{12\ldots j+1}+n_j-z_{j}} \prod_{\substack{j\leq l\\i \notin [j,l]}}\left( 1-\prod_{k=j}^l w_k \right)^{\tilde{s}_{j+1,l+2}}\nonumber \\
    &\times \prod_{\substack{j\leq l\\i \in [j,l]}}\left(  \sum_{a_{jl}=0}^\infty \binom{\tilde{s}_{j+1,l+2}}{a_{jl}} (-1)^{a_{jl}} \left(\prod_{\substack{k=j\\k\neq i}}^l w_k  \right)^{a_{jl}} \right) \nonumber \\
    &\times \int_0^1 dw_i \ w_i^{s_{12\ldots i+1}+n_i+\sum_{j\leq l}a_{jl}-z_i} \label{deformed hypergeometric2}
\end{align}
Notice that we factor out the variable $w_i$ from the others in the last line. Integrating out the $w_i$-variable turn (\ref{deformed hypergeometric2}) into
\begin{align}
    B_n(\{ n\},z)=\sum_m \frac{\mathcal{B}_i(\{n\},m,z)}{z_i-(s_{12\ldots i+1}+n_i+m+1)}
\end{align}
where 
\begin{align}
    \mathcal{B}_i(\{n\},m,z)=&\left(\prod_{\substack{k=1\\k\neq i}}^{n-3}\int_0^1 dw_k \right) \prod_{\substack{j=1\\j\neq i}}^{n-3} w_j^{s_{12\ldots j+1}+n_j-z_{j}} \prod_{\substack{j\leq l\\i \notin [j,l]}}\left( 1-\prod_{k=j}^l w_k \right)^{\tilde{s}_{j+1,l+2}} \nonumber \\
    &\times (-1)^{m+1}\prod_{\substack{j\leq l\\i \in [j,l]\\(j,l)\neq (1,n-3)}}\left(  \sum_{a_{jl}=0}^\infty \binom{\tilde{s}_{j+1,l+2}}{a_{jl}}  \left(\prod_{\substack{k=j\\k\neq i}}^l w_k  \right)^{a_{jl}}\right) \nonumber \\
    &\times \binom{\tilde{s}_{2,n-1}}{m-\sum_{(j,l)\neq (1,n-3)} a_{jl}} \left(\prod_{k=1}^{n-3}w_k\right)^{m-\sum_{(j,l)\neq (1,n-3)}a_{jl}}. \label{function Bi}
\end{align}
We denote the index $m$ as sum of all indices $a_{jl}$
\begin{equation}
    m=\sum_{\substack{j\leq l\\i \in [j,l]}} a_{jl}.
\end{equation}
Accordingly, we need to write one index out of all the $a_{ij}$'s in terms of $m$. Our choice is $a_{1,n-3}$ as this index always appears in the formulation regardless of the index $i$ we consider. Remember that when $i=1$, the expression for 
$\mathcal{B}_1(\{n\},m,z)$ matches with that presented in (\ref{B1-2}).

Now, one can obtain the residue
\begin{align}
    \sum_{z_{\text{pole}}} \text{Res}_{z=z_{\text{pole}}}\bigg(\frac{B_n(\{ n\},z)}{z}\bigg)=\sum_{i=1}^{n-3}\sum_{m=0}^\infty \frac{\mathcal{B}_i(\{n\},m,z^*_{im})}{s_{12\ldots i+1}+n_i+m+1} \label{residue zi}
\end{align}
where 
\begin{equation}
   z^*_{im}=\frac{s_{12\ldots i+1}+n_i+m+1}{2\alpha' q\cdot \sum_{j=2}^{i+1}k_j}.
\end{equation}
Consequently, the general expression for color-ordered open string amplitude is then written as 
\begin{align}
     \mathcal{A}(1,2,3,\ldots,n)=\sum_{i=1}^{n-3}\sum_{m=0}^\infty \sum_I \mathcal{K}_I(z^*_{im})\frac{\mathcal{B}_i(\{n^I\},m,z^*_{im})}{s_{12\ldots i+1}+n_i+m+1}. \label{final BCFW open}
\end{align}
Note that the function $\mathcal{K}_I$ is also evaluated at $z^*_{im}$ as well since the function is $z$-dependent when the momenta are shifted. 

It is clear that the denominators in the expression (\ref{final BCFW open}) correspond to propagators of the on-shell intermediate string states.  For a specific example of $n$-tachyon scattering, $n_i=-2$. The denominators suggest the full spectrum of open strings, i.e.
\begin{equation}
    (k_1+k_2+\ldots+k_{i+1})^2+\frac{(m-1)}{\alpha'}
\end{equation}
for a non-negative integer $m$. In the case of $n-1$ tachyons and one gauge boson scattering assigning polarization for the vector boson as $\xi_1$, one obtains $\tilde{n}_{1i}=-1$ for $i=2,3,\ldots, n$ through linearizing the $\xi_1$ in $\mathcal{F}_n$. This also gives $n_i=-2$ providing the same mass spectrum of propagating open strings. On the other hand, one can use the requirement of the intermediate states to be on-shell to inversely identify all $\tilde{n}_{ij}$  from the function $\mathcal{F}_n$ via (\ref{n n}).

The existence of scattering channel $s_{12\ldots i+1}$ implies that the function (\ref{function Bi}) can be factorized into two objects referring to the amplitudes of lower points. For the channel $s_{12\ldots i+1}$, one can write
\begin{align}
    \mathcal{B}_i(\{n\},m,z)=&\prod_{\substack{j\leq l\\i \in [j,l]\\(j,l)\neq (1,n-3)}}\left(  \sum_{a_{jl}=0}^\infty \binom{\tilde{s}_{j+1,l+2}}{a_{jl}} \right) \binom{\tilde{s}_{2,n-1}}{m-\sum_{(j,l)\neq (1,n-3)} a_{jl}} \nonumber \\
    &\times B_{i,n}(\{n\},\{a\},m,z) \times \widetilde{B}_{i,n}(\{n\},\{a\},m,z)
\end{align}
where
\begin{align}
    B_{i,n}(\{n\},\{a\},m,z)=&\left(\prod_{j=1}^{i-1}\int_0^1 dw_j \right) \prod_{k=1}^{i-1} w_k^{s_{12\ldots k+1}+n_k-z_k+m-a_{ii}-\sum\limits_{\small{\substack{j<l \\ j,l=k+1}}}^{n-3} a_{jl}} \nonumber \\
    &\times \prod_{l=k}^{i-1}\left( 1-\prod_{r=k}^l w_r \right)^{s_{k+1,l+2}+n_{k+1,l+2}} 
\end{align}
and 
\begin{align}
    \widetilde{B}_{i,n}(\{n\},\{a\},m,z)=&\left(\prod_{j=i+1}^{n-3}\int_0^1 dw_j \right) \prod_{k=i+1}^{n-3} w_k^{s_{12\ldots k+1}+n_k-z_k+m-a_{ii}-\sum\limits_{\small{\substack{j<l \\ j,l=1}}}^{k-1} a_{jl}} \nonumber \\
    &\times \prod_{l=k}^{n-3}\left( 1-\prod_{r=k}^l w_r \right)^{s_{k+1,l+2}+n_{k+1,l+2}}.
\end{align}
The functions $B_{i,n}(\{n\},\{a\},m,z)$ and $\widetilde{B}_{i,n}(\{n\},\{a\},m,z)$ are in the forms of open string amplitudes with $i+1$ and $n-i$ external legs respectively. This signifies the fact that the on-shell recursion relations boil down a scattering amplitude in terms of lower-point amplitudes.

\section{Conclusions}

In conclusion, we have developed a systematic approach to derive on-shell recursion relations for open string amplitudes. The derivation, which is based on the same technique initiated by Britto et al. \cite{Britto_2005, Britto2005NewRR}, starts from applying the BCFW shift to an open string amplitude written in terms of multiple Gaussian hypergeometric functions. The shift generates simple poles corresponding to scattering channels of intermediate states. Using the residue theorem, we expressed a general expression for $n-$point open string on-shell recursion relations as in (\ref{final BCFW open}).



\bibliographystyle{elsarticle-num} 
\bibliography{example}


\end{document}